\documentclass[aps,prd,twocolumn,showpacs,preprintnumbers,amsmath,amssymb]{revtex4-1}

\usepackage{graphicx}
\usepackage[usenames,dvipsnames]{xcolor}
\usepackage{braket}
\usepackage{amssymb}
\usepackage{amsfonts}
\usepackage{amsmath}
\usepackage{bbm} 
\usepackage{bbold}
\usepackage{siunitx}
\usepackage[colorlinks]{hyperref}
\usepackage{makeidx}
\definecolor{winered}{rgb}{0.8,0,0}
\definecolor{darkb}{rgb}{0,0,0.8}
 \usepackage{hyperref}
\hypersetup{
    colorlinks=true,
    citecolor=blue,
    linkcolor=winered,
    filecolor=red,      
    urlcolor=darkb,
}
 
\usepackage{lineno}
  \usepackage{young}
\begin{document}
\title{A linear sigma model for multiflavor gauge theories}
\author{Y. Meurice$^1$}
\affiliation{$^1$ Department of Physics and Astronomy, The University of Iowa, Iowa City, IA 52242, USA }
\definecolor{burnt}{cmyk}{0.2,0.8,1,0}
\def\lt{\lambda ^t}
\def\note{note}
\def\beq{\begin{equation}}
\def\enq{\end{equation}}
\newcommand{\Tr}{\text{Tr}}

\def\gf{U(N_f)_L\bigotimes U(N_f)_R}
\def\ls{\lambda_\sigma}
\def\la{\lambda_{a0}}
\def\pdp{\phi^\dagger\phi}
\def\bpi{\boldsymbol\pi}
\def\bao{\bf a_0}
\def\etp{\eta'}
\date{\today}
\begin{abstract}
We consider a linear 
sigma model describing $2N_f^2$  bosons ($\sigma$, $\bao$, $\etp$ and $\bpi$) as an approximate effective theory for a $SU(3)$ local gauge theory with $N_f$ Dirac fermions in the fundamental representation.  
The model has a renormalizable $\gf$ invariant part, which has an approximate $O(2N_f^2)$ symmetry, and two additional terms, one describing the effects of a  $SU(N_f)_V$ invariant
mass term and the other the effects of the axial anomaly. We calculate 
the spectrum for arbitrary $N_f$. Using preliminary  and published lattice results from the LatKMI collaboration, we found combinations of the masses that vary slowly with the explicit chiral symmetry breaking and $N_f$. This suggests that the anomaly term 
plays a leading role in the mass spectrum and that simple formulas such as $M_\sigma^2\simeq (2/N_f-C_\sigma)M_{\etp}^2$ should apply in the chiral limit. Lattice measurements of $M_{\etp}^2$ and of approximate constants such as $C_\sigma$ could help locating the boundary of the conformal window.
We show that our calculation can be adapted  for arbitrary representations of the gauge group and in particular to the minimal model with two sextets, where similar patterns are likely to apply.
 \end{abstract}

\maketitle
\section{Introduction}
The linear version of the sigma models introduced by Gell-Mann and Levy \cite{sigma} has played an influential role \cite{benlee} in the establishment of the standard model. In today's usage \cite{Scherer:2005ri,bijnensreview}, the nonlinear versions not involving the 
$\sigma$-particle ($f_0(500)$) are favored in quantum chromodynamics (QCD) based low-energy calculations.  However, when dealing with the explicit breaking of the axial $U(1)_A$ symmetry, 
linear models are used to describe the $\eta'$ \cite{PhysRevD.3.2874,PhysRevD.21.3388,tHooftphysrep,Meuricea0}. In addition to the $\eta '$ and $\bpi$, the linear models involve the $\sigma$ and the $\bao$ ($0^+$ isovectors, $a_0(980)$). In QCD, the $\sigma$ and the $\bao$ are correctly considered as ``heavy" particles compared to the ``light" pions which - unlike their $0^+$ counterparts - become massless in the chiral limit. 
However if enough light flavors are added, the separation between light and heavy changes, and the possibility of 
having a light $\sigma$ is quite attractive from the low energy point of view.  

The spectrum of multiflavor gauge theories has been vigorously investigated in the context of finding hypothetical strongly interacting particles responsible for the formation of the Brout-Englert-Higgs particle 
and the electroweak symmetry breaking. For recent reviews of the physics motivations and the literature on this subject, we recommend Refs. \cite{bsr,dgrmp,nogrev}. 
The estimation of the mass of flavor singlets using lattice simulations involves disconnected diagrams and is 
computationally expensive. There are only a few available results. For instance, the estimation of the $\eta'$ in QCD has only be achieved recently \cite{Christ:2010dd,Gregory:2011sg,Ottnad:2015hva}. Similarly, there are only few results available in the multiflavor case. 

Recently, light $\sigma$ masses were found for $SU(3)$ gauge theories with 
8 \cite{Aoki:2014oha,Appelquist:2016viq,Aoki:2016wnc,Gasbarro:2017fmi} and 12 \cite{Aoki:2013zsa,Aoki:2016wnc} fundamental flavors and also for 2 sextets \cite{Fodor:2015vwa}. 
In addition, preliminary results \cite{latkmi16,latkmi17} concerning the mass of the $\etp$ were announced at recent conferences. With this information, we would like to investigate the possibility that the explicit breaking
of the axial $U(1)_A$ symmetry, which depends in a distinct way on $N_f$, plays an important role in the determination of the spectrum and the boundary of the conformal window where a nontrivial infrared fixed point is present \cite{bz,st05,ds07}. 
As the models mentioned above have a low energy behavior significantly different from QCD, it is very desirable from the point of view of model building to have a simple effective description of this behavior.

In this paper, we consider a generalization of the models discussed for $N_f=2$  \cite{tHooftphysrep}, and  $N_f=3$ \cite{PhysRevD.3.2874,PhysRevD.21.3388,Meuricea0} in the context of QCD. This is a linear 
sigma model describing $2N_f^2$  bosons ($\sigma$, $\bao$, $\etp$ and $\bpi$), using the QCD terminology. We will use  it here 
as an approximate effective theory for a $SU(3)$ local gauge theory with $N_f$ Dirac fermions in the fundamental representation.  The Lagrangian  has a renormalizable $\gf$ invariant part and two additional terms, one representing a $SU(N_f)_V$ invariant
mass term and the other the axial anomaly. 

A $SU(N_f)_V$-invariant effective theory for pions is discussed in \cite{Bijnens:2009qm} for arbitrary $N_f$. 
There has been a recent interest to include the $\sigma$ in dilatonic effective theories \cite{PhysRevD.94.054502,Golterman:2016hlz,Appelquist:2017wcg}. 
We would like to briefly motivate the inclusion of the ``complex partners" the $\etp$ and $\bao$. In the case of $N_f$=2, the term 
that we use to describe the axial anomaly effect is simply a mass term with alternate signs:
\beq
V_a|_{N_f=2}\propto(\etp^2-\sigma^2+\bao^2-\bpi^2).
\label{eq:n2}
\enq
The $N_f=2$ model was used by 't Hooft \cite{tHooftphysrep} to explain the role that the instantons play in the spectrum because if we replace the effective bosonic 
degrees of freedom in (\ref{eq:n2}) by their quark content (${\bao}  \sim  \bar{\psi} \bf {\tau} \psi$ etc ..),  we recognize a term of the 't Hooft determinant \cite{thooft76}. In the following, we discuss the generalization for an arbitrary number of flavors $N_f$. As we will see, the axial anomaly term is essential to get a light $\sigma$.

The linear sigma model is presented in Sec. \ref{sec:model}. The tree level spectrum is calculated for arbitrary $N_f$ in Sec. \ref{sec:spectrum}. In Sec. \ref{sec:ratios}, we introduce dimensionless quantities involving the masses. 
Preliminary  \cite{latkmi17} and published \cite{Aoki:2016wnc} lattice results, indicate that they vary slowly with the explicit chiral symmetry breaking and $N_f$. This provides approximate mass formulas that, if  properly refined 
by future lattice calculations may help identify instabilities for large enough $N_f$ and pinpoint the boundary of the conformal window. 
In Sec. \ref{sec:higher} we show how to extend our results for fermions in an arbitrary representation of the gauge group. In the conclusions, we discuss the relevance of the results for future lattice calculations. 

\section{The model}
\label{sec:model}
Following Refs. \cite{PhysRevD.3.2874,PhysRevD.21.3388,tHooftphysrep,Meuricea0}, 
we consider a $N_f\times N_f$ matrix of effective fields $\phi_{ij}$  having the same quantum numbers as  $\bar{\psi}_{Rj}\psi_{Li}$ with the summation over the color indices implicit. 
Under a general transformation of $\gf$, we have 
\beq
\label{eq:phit}
\phi \rightarrow U_L\phi U_R^{\dagger} \ .
\enq
We now use a basis of $N_f\times N_f$ Hermitian matrices $\Gamma^\alpha$ such that 
\beq
\label{eq:norm}
Tr(\Gamma^\alpha\Gamma^\beta)=(1/2)\delta^{\alpha\beta}, 
\enq
to express $\phi$ in terms of $N_f^2$ scalars ($0^+$ in $J^P$ notation), denoted $S_\alpha$, and $N_f^2$ pseudoscalars ($0^-$), denoted $P_\alpha$:
\beq
\label{eq:phip}
\phi=(S_\alpha+iP_\alpha)\Gamma^\alpha, 
\enq
with a summation over $\alpha=0,1,\dots N_f^2-1$. We use the convention that $\Gamma_0=\mathbb{1}/\sqrt{2N_f}$ while the remaining $N_f^2-1$ matrices are traceless. 

We introduce the diagonal subgroup $U(N_f)_V$ defined by the elements of $\gf$ such that $U_L=U_R$. Using Eqs. (\ref{eq:phip}) and 
(\ref{eq:phit}) we see that under $U(N_f)_V$, $S_0$ and $P_0$ are singlets denoted $\sigma$ and $\etp$ respectively while the remaining components transform like the adjoint representation and are denoted ${\bao}$ and $\bpi$ respectively. 

We consider the effective Lagrangian 
\beq
\label{eq:lag}
{\mathcal{L}}=Tr\partial_\mu\phi\partial^\mu\phi^\dagger-V
\enq
with the potential split into three parts
\beq
V=V_0+V_a+V_m,
\enq
that we now proceed to define and discuss.
The first term is  the most general $\gf$ invariant renormalizable expression: 
\begin{eqnarray}
\label{eq:vo}
V_0&\equiv&-\mu^2Tr(\phi^\dagger\phi)+(1/2)(\ls-\la)(Tr(\pdp))^2\nonumber\\
&\ &+(N_f/2)\la Tr((\pdp)^2). 
\end{eqnarray}
The use of $\ls-\la$ will become clear when we write the mass formulas. 
The stability of $V_0$ is discussed in Sec. \ref{sec:ratios}. 
Note that first two terms and the kinetic term have a larger group of symmetry $O(2N_f^2)$. 
The second term 
\beq
V_a\equiv-2(2N_f)^{N_f/2-2}X(det\phi + det\phi^\dagger), 
\enq
$\   $
\vskip3pt
\noindent
is invariant under $SU(N_f)_L\bigotimes SU(N_f)_R$ but breaks the axial $U(1)_A$. It takes into account the effect of the axial anomaly for the fundamental representation. The generalization to arbitrary representations is discussed in Sec. \ref{sec:higher}. The prefactor $2(2N_f)^{N_f/2-2}$ is chosen in order to make the expression of the spectrum as simple as possible. The parameter $X$ has a mass dimension $4-N_f$. Related effective descriptions of the 
breaking of the $U(1)_A$ can be found in the literature \cite{DiVecchia:1980yfw,Kawarabayashi:1980dp,Ohta:1981ai,Kawarabayashi:1980uh}.

Finally the third term represent the effect of mass term which is the same for the $N_f $ flavors: 
\beq
V_m\equiv-(b/\sqrt{2N_f})(Tr\phi+Tr\phi^\dagger)=-b\sigma.
\enq
It is invariant under $SU(N)_V$. 

In the following, we assume that chiral symmetry is spontaneously broken by a 
$SU(N_f)$-invariant vacuum expectation value (v.e.v.): 
\beq
\label{eq:vev}
\left\langle\phi_{ij}\right\rangle=v\delta_{ij}/\sqrt{2N_f}.
\enq
This amounts to say that $\left\langle \sigma \right\rangle=v$ while the other v.e.v.s are zero. 
We impose that 
\beq
\partial V/\partial \phi |_{\left\langle\phi \right\rangle} =0. 
\enq
Thanks to the simple form of the v.e.v.s in Eq. (\ref{eq:vev}), these $N_f^2$ equations reduce to a single one:
\beq
\label{eq:min}
-\mu^2v+(1/2)\ls v^3 -(X/N_f)v^{N_f-1}=b.
\enq
\section{The spectrum}
\label{sec:spectrum}
We can now calculate the tree level spectrum. 
The normalization  (\ref{eq:norm}) implies that the  kinetic term in Eq. (\ref{eq:lag}) is canonical: 
\beq
\label{eq:kin}
Tr\partial_\mu\phi\partial^\mu\phi^\dagger=(1/2)\sum_\alpha(\partial_\mu S_\alpha \partial^\mu S_\alpha+\partial_\mu P_\alpha \partial^\mu P_\alpha).
\enq
The  mass of the fields are then obtained as the second derivatives in an obvious way (as for free Klein-Gordon fields). 
In addition, the unbroken $SU(N_f)_V$ symmetry simplifies the formulas which can be expressed in terms of 4 masses:

\begin{eqnarray}
\partial^2 V/\partial S_0\partial S_0 |_{\left\langle\phi \right\rangle}&=&M^2_\sigma\nonumber\\
\partial^2 V/\partial S_i\partial S_j |_{\left\langle\phi \right\rangle}&=&\delta_{ij}M^2_{a0} \nonumber\\
\partial^2 V/\partial P_0\partial P_0 |_{\left\langle\phi \right\rangle}&=&M^2_{\etp} \nonumber\\
\partial^2 V/\partial P_i\partial P_j |_{\left\langle\phi \right\rangle}&=&\delta_{ij}M^2_\pi.
\end{eqnarray}
By convention, in the 4 above equations,  the latin indices run from 1 to $N_f^2-1$ (isovector indices in the $N_f=2$ language). 
The notation $M_\sigma^2$ does not mean that this quantity is automatically positive. A negative value could indicate an instability. 
Since ${\left\langle\phi \right\rangle}$ is proportional to the identity, the second derivatives can be easily calculated at the assumed v.e.v.s.
For instance, the derivative of the determinant involves the inverse which can then be easily evaluated. This would not be the case, for an arbitrary breaking where we would need to use $f$ and $d$ symbols \cite{Meuricea0}. 
For the pions, we have:
\beq
M^2 _\pi=-\mu^2+(1/2)\ls v^2 -(X/N_f)v^{N_f-2}.
\enq
Using the minimization condition (\ref{eq:min}), this  can be recast in the form
\beq
M_\pi^2v=b.
\enq
In other words in absence of the explicit mass breaking ($b=0$), we have the familiar result $M_\pi^2v=0$ and $v\neq0$ implies 
that in this chiral limit, the pions are exactly massless Nambu Goldstone bosons. The v.e.v. $v$ is related to the pion decay constant in the following way:
\beq
f_\pi=\sqrt{2/N_f} v .
\enq
This result can be obtained by considering an $U(N_f)_A$ transformation:
\beq
\phi\rightarrow U\phi U\simeq \phi +i\omega_\alpha\{ \Gamma^\alpha,\phi\}, 
\enq
and showing that the Noether current satisfies the PCAC relation 
\beq
\partial ^\mu J_\mu^\alpha=b\sqrt{2/N_f}P^\alpha.
\enq
We used the convention that 
\beq
\bra{\Omega}J_\mu^\alpha(x)\ket{P^\beta(p)}=i\delta^{\alpha \beta}f_\pi p_\mu{\rm e}^{-ipx}.
\enq

The other results for the spectrum can be written in a compact way:
\begin{eqnarray}
\label{eq:main}
M_{\etp}^2-M_\pi^2&=&Xv^{N_f-2}\nonumber\\
M_\sigma^2-M_\pi^2&=&\ls v^2-(1-2/N_f)Xv^{N_f-2}\\
M_{a0}^2-M_\pi^2&=&\la v^2+(2/N_f)Xv^{N_f-2}\nonumber .
\end{eqnarray}
One can check that these results agree with the corresponding results \cite{Meuricea0} for $N_f=3$ in the $SU(3)_V$ limit. 
In the chiral limit ($b=0$), Eqs. (\ref{eq:main}) reduce to 
\begin{eqnarray}
\label{eq:chiral}
M_\sigma^2&=&\ls v^2-(1-2/N_f)M_{\etp}^2 \\
M_{a0}^2&=&\la v^2+(2/N_f)M_{\etp}^2 \nonumber .
\end{eqnarray}
The sign of the interaction in the anomaly term $V_a$ has been chosen in such a way that 
$M_{\etp}^2\geq M_\pi^2$ as in QCD. This feature persists in more general situations. 
This implies that for $N_f\geq3$ the mass of the $\sigma$ has two contributions of opposite sign. 
In order to have $M_\sigma \simeq M_\pi$, the two contributions should either be small separately or cancel each other. 
We will see that the second possibility seems to be realized in a certain number of situations. 

Notice that if $b$, $\la$ and $X$  are set to zero, $M_{\etp}=M_{a0}=0$ and the $\etp$ and $\bao$ could be interpreted as $N_f^2$ additional Nambu-Goldstone bosons. This is because 
in this limit, the effective Lagrangian has an $O(2N_f^2)$ symmetry and 
the v.e.v. of the $\sigma$ breaks it down to $O(2N_f^2-1)$ resulting in a total of $2N_f^2-1$ Nambu-Goldstone bosons.

In the next section, we explain that: 1) it is legitimate to consider the $\la v^2$ contribution as small, 2) the $M_{\etp}^2$ is a significant contribution 
partially suppressed by the $1/N_f$ factor in the $M_{a_0}^2$ formula. In that sense, it seems legitimate to treat the $\bao$ as  light particles in
some region of the ($m_f$,$N_f$) plane. We should add that the chiral limit obtained from a linear fit of the data for $M_{a_0}$ (see Fig. 28 of Ref. \cite{Aoki:2016wnc}) gives a value significantly smaller than masses quoted with a finite $m_f$. For instance, at $am_f=0.03$, we have $aM_{a_0}\simeq0.46$, while the chiral extrapolation is $aM_{a_0}\simeq0.16$. In other words the slope in the $aM_{a_0}$ versus $am_f$ graph is rather large (about 10). 

\section{Dimensionless ratios}
\label{sec:ratios}

In order to allow comparisons of numerical results at different lattice spacings, we introduce the dimensionless ratios:
\beq
R_{\sigma}\equiv \ls v^2/M_{\etp}^2,
\enq
and 
 \beq
R_{a_0}\equiv \la v^2/M_{\etp}^2.
\enq
We want to test the idea that these quantities vary slowly with the explicit breaking of chiral symmetry (due to the mass of the fermions $m_f$) and $N_f$. 
If it is the case, the mass formulas have simple approximate form which could provide a nice intuitive picture. 
To make things completely clear, the ratios should be understood as functions of the spectroscopic data, namely
\begin{eqnarray}
R_{\sigma}&=&(M_\sigma^2-M_\pi^2)/M_{\etp}^2+(1-2/N_f)(1-M_\pi^2/M_{\etp}^2) \nonumber \\
R_{a_0}&=&(M_{a_0}^2-M_\pi^2)/M_{\etp}^2 -(2/N_f)(1-M_\pi^2/M_{\etp}^2).
\end{eqnarray}
\begin{figure}[hh]
\includegraphics[width=3.in]{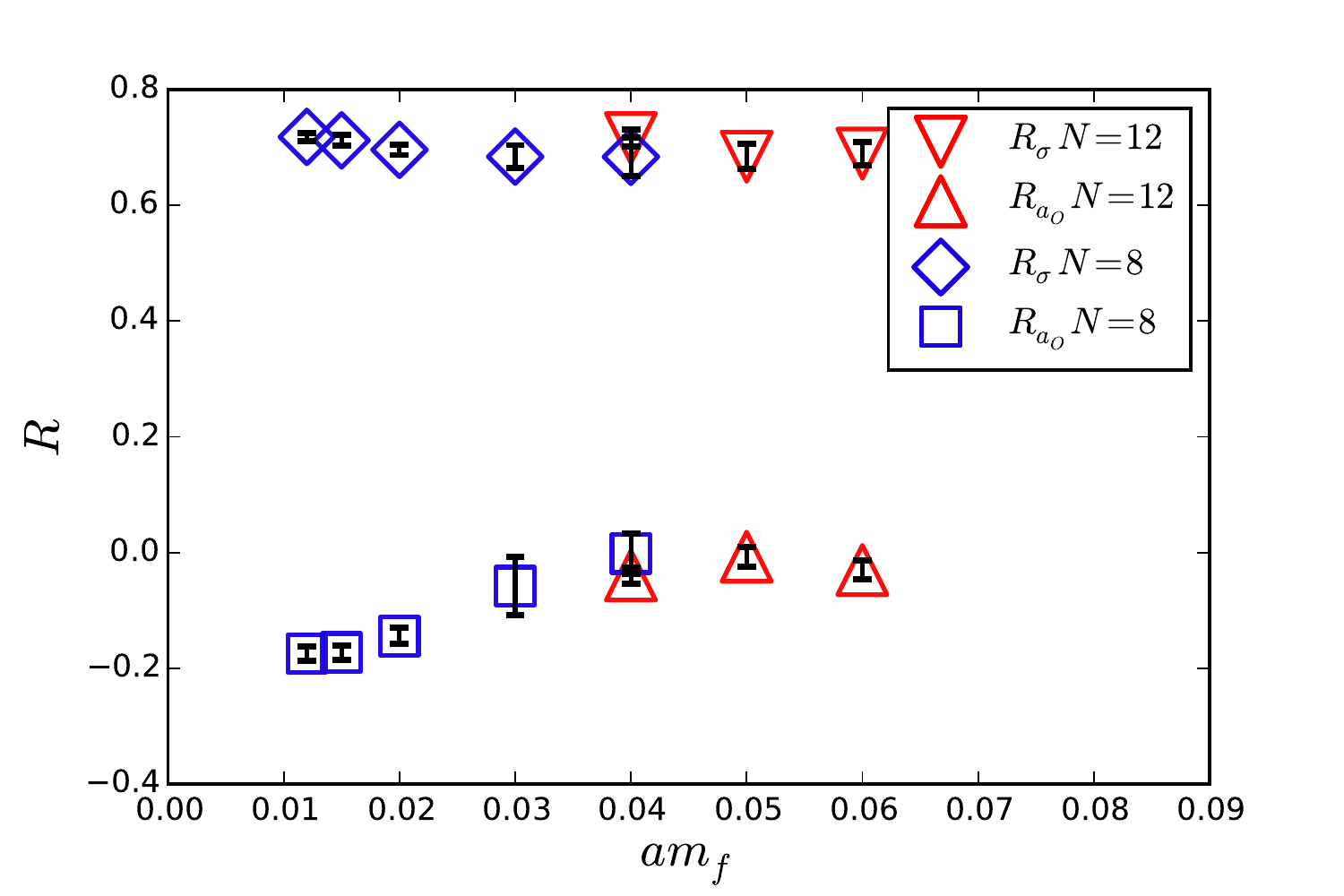}
\caption{\label{fig:rat}
$R_{\sigma}$ for $N_f$=8 (diamonds) and 12 (upside-down triangles) and $R_{a_0}$ for $N_f$=8 (squares)  and 12 (triangles), versus $am_f$.}
\end{figure}

\begin{figure}[hh]
\includegraphics[width=3.in]{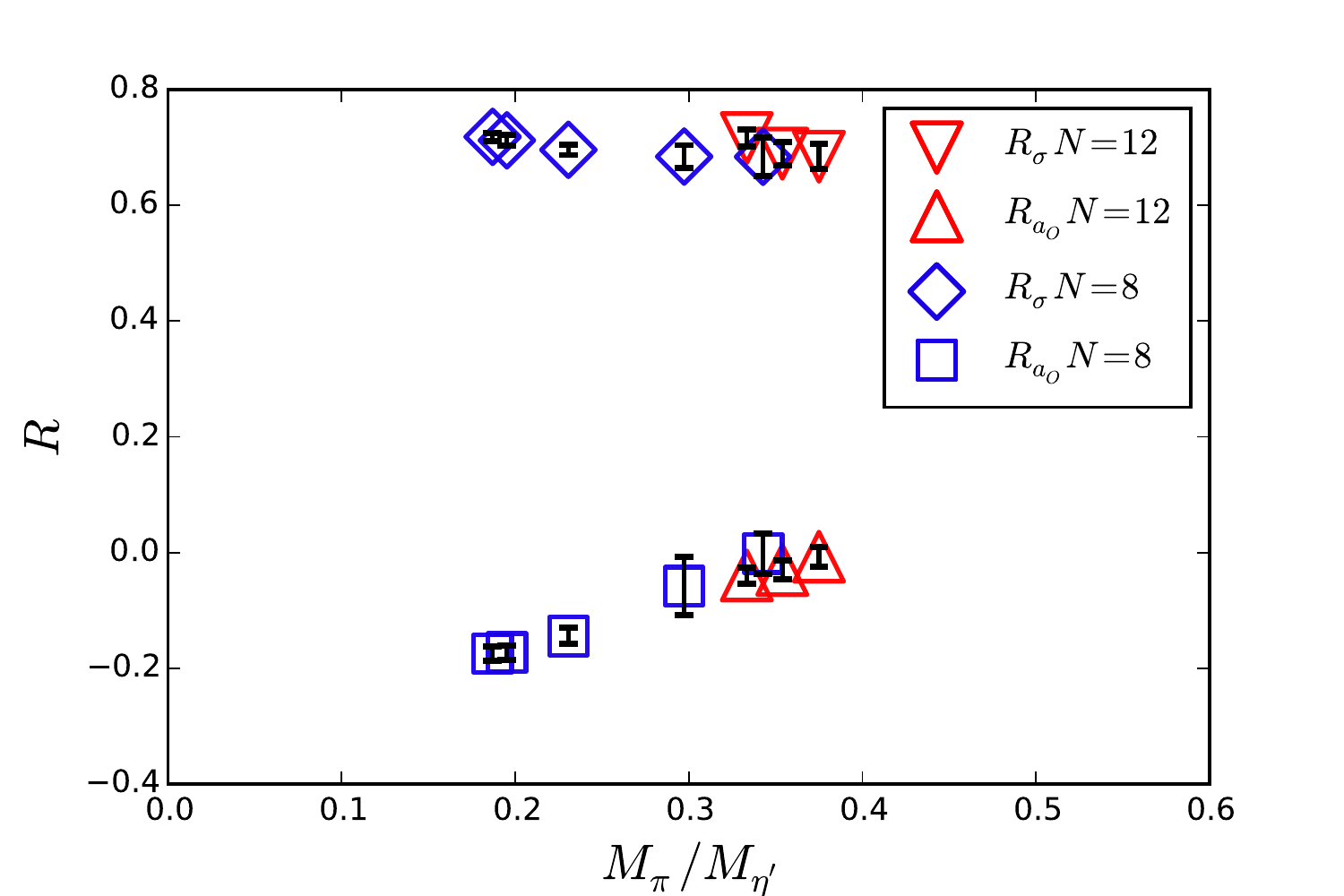}
\caption{\label{fig:ratvrat}
$R_{\sigma}$ for $N_f$=8 (diamonds) and 12 (upside-down triangles) and $R_{a_0}$ for $N_f$=8 (squares)  and 12 (triangles), versus 
$M_\pi/M_{\etp}$.}
\end{figure}

Using the preliminary results presented at Lattice 2017 \cite{latkmi17} and the published  results of Ref. \cite{Aoki:2016wnc} we obtained Fig. \ref{fig:rat}.
As explained in the Appendix, these results have been obtained by using the largest volume available for each mass. The error bars do not take into account 
the finite volume effects and could be significantly larger if we had taken them into account. 
Nevertheless, Fig. \ref{fig:rat} is consistent with the idea of slowly varying ratios. We propose the 
order of magnitude estimates:
\beq
R_{\sigma}\simeq 0.7 +\delta_{\sigma}(m_f,N_f),
\label{eq:ras}
\enq
and 
 \beq
R_{a_0}\simeq -0.1+\delta_{a_0}(m_f,N_f),
\label{eq:rao}
\enq
where the $\delta(m_f,N_f)$ describe slow variations that need to be studied carefully in the future. 
The spectroscopic data is often given in terms of $m_f$ in lattice spacing units, but we could as well parametrize the residuals in 
terms of another symmetry breaking parameter for instance $M_\pi^2$. Another possibility is to use the dimensionless ratio $M_\pi/M_{\etp}$ 
which does not involve the lattice spacing and may be more suitable to compare results at different $N_f$. This is done in Fig. \ref{fig:ratvrat}.

Fig. \ref{fig:rat} shows that for small $m_f$, we have negative values of $R_{a_0}$. Can $\la$ be negative? 
The stability of the effective potential requires that $V_0$ defined in (\ref{eq:vo}) stays positive for large values of the field $\phi$. 
Using the symmetry $\gf$ of $V_0$, we can diagonalize $\phi$. $\phi^\dagger \phi$ is then a diagonal matrix with positive diagonal terms $|\alpha_i|^2$ and the sum of the two quartic terms of $V_0$ will remain positive when $\la=-|\la|$ is negative provided that 
\beq
(1/2)(\ls+|\la|)(\sum_{i=1}^{N_f} |\alpha_i|^2)^2 \geq (N_f/2)|\la |\sum_{i=1}^{N_f}|\alpha_i|^4.
\enq
This inequality should remain valid for any choice of $\alpha_i$. Considering the case where only one $|\alpha_i|$ becomes arbitrarily large, we get the requirement. 
\beq
\label{eq:req}
\ls \geq(N_f-1)|\la | .
\enq
Using the inequality
\beq
(\sum_{i=1}^{N_f} |\alpha_i|^2)^2\geq \sum_{i=1}^{N_f}|\alpha_i|^4, 
\enq
we see that the requirement (\ref{eq:req}) is also a  sufficient condition in general. 

Using the estimates of Eqs. (\ref{eq:ras}) and (\ref{eq:rao}) in the chiral limit, 
we obtain the simple approximate picture: 
\begin{eqnarray}
\label{eq:sig}
(M_\sigma/M_{\etp})^2&\simeq &2/N_f-0.3+\delta_{\sigma}(0,N_f),\\
\label{eq:az}
(M_{a_0}/M_{\etp})^2&\simeq &2/N_f-0.1
+\delta_{a_0}(0,N_f).
\end{eqnarray}
The stability bound  $M_{\sigma}^2\geq 0$ 
could provide an estimate of $N_{fc}$.
The idea of having $R_{a_0}<0$ is attractive because it implies additional stability bounds on $N_f$ coming from $M_{a_0}^2\geq 0$. For instance, if we use the rough chiral limit estimates $R_{\sigma}\approx 0.8$ and $R_{a_0}\approx -0.2$, suggested by Fig. \ref{fig:rat}, we obtain the same $N_{fc} \approx 10$  from (\ref{eq:sig}) with $\delta_{\sigma}(0,N_f)=0.1$, and (\ref{eq:az})  with $\delta_{a_0}(0,N_f)=-0.1$, respectively. In addition, the stability 
 bound of Eq. (\ref{eq:req}) implies
\beq
\label{eq:sb}
N_f\leq 1+R_{\sigma}/|R_{a_0}|.
\enq
With our chiral limit estimates this corresponds to $N_{fc} \approx 5$. 
These numbers should not be taken too seriously. They are just meant to illustrate the fact that a careful study of the residuals $\delta(m_f,N_f)$ might help to pinpoint the boundary of the conformal window.

Fig. \ref{fig:rat} suggests that we could try to find functional relations that are approximately $N_f$-independent, at least for sufficiently massive theories.
For the common mass $am_f=0.04$, the ratios are almost identical despite significantly different meson masses (see Table I in the Appendix). 
Results at smaller $am_f$ would be interesting to see if the chiral limits are very different for $N_f=12$ and $N_f=8$, as expected if they are on opposite sides of the conformal window. Since $am_f$ depend explicitly on the lattice spacing $a$, we have also plotted the ratios versus the dimensionless ratio $M_\pi/M_{\etp}$ in Fig. \ref{fig:ratvrat}.  
This may be a better way to proceed with the functional relations idea. We also have one data point available \cite{latkmi17} for $N_f=4$ at much smaller volume 
and $am_f=0.01$ providing  $R_\sigma\sim 0.55$ which is close to 0.7. More data for this case as well as $N_f$= 6 and 10 would be very desirable to study the residual functions $\delta$.

\section{Modification for higher representations}
\label{sec:higher}
If the microscopic theory is defined by fermions in higher dimensional representations, for instance $SU(3)$ sextets (the twice symmetric 
representation), we need to modify the axial anomaly term as
\beq
V_a(K)=-2(2N_f)^{N_f/2-2}X((det\phi)^K + (det\phi^\dagger)^K), 
\enq
with $K$ is the number of zero modes in an instanton background. It also appears in the one-loop coefficient of the Callan-Symanzik beta function. It can be written as  $K=2T[R]$ where $T[R]$ is the trace normalization of the  
representation $R$ (see Ref. \cite{Fodor:2009nh} for a lattice discussion).
For instance, for $SU(3)$ sextets, $K=5$. 
In general, $T[R]$ can be calculated  in terms of the Casimir operator of the representation $C_2(R)$ using the relation $T[R]=d(R)C_2(R)/d(Adjoint)$. The coupling $X$ has now a dimension $4-KN_f$. 

The minimization equation (\ref{eq:min}), $M_\pi^2$ and $M_{a_0^2}$ can be generalized by changing 
\beq
X\rightarrow \tilde{X}\equiv XK\bigg(v/\sqrt{2N_f}\bigg)^{N_f(K-1)},
\enq
into the equations for the fundamental representation derived above. 
For $M_{\etp}^2$ and $M_{\sigma}^2$, in addition of this substitution we need to add a term $(K-1)\tilde{X}$ with a positive sign for  $\etp$ and a negative sign for $\sigma$. In the ``minimal" case of $N_f=2$ $SU(3)$ sextets studied in Refs. \cite{Dietrich:2005jn,DeGrand:2008kx,DeGrand:2012yq,Fodor:2012ty,Fodor:2015vwa}:
\begin{eqnarray}
\label{eq:main6}
M_{\etp}^2-M_\pi^2&=&(25/256)Xv^8\nonumber\\
M_\sigma^2-M_\pi^2&=&\ls v^2-(5/64)Xv^8\\
M_{a0}^2-M_\pi^2&=&\la v^2+(5/256)Xv^8\nonumber .
\end{eqnarray}
Again we see that the $M^2_{\sigma}$ receives contributions of opposite signs, making $M_{\sigma}^2<M_\pi^2$ observed in Ref. \cite{Fodor:2015vwa} 
possible in our model.
In the chiral limit ($b=0$), Eqs. (\ref{eq:main6}) reduce to 
\begin{eqnarray}
\label{eq:chiral6}
M_\sigma^2&=&\ls v^2-(4/5)M_{\etp}^2 \\
M_{a0}^2&=&\la v^2+(1/5)M_{\etp}^2 \nonumber .
\end{eqnarray}
Calculating $M_{\etp}^2$ for this model would be quite interesting. 
\section{Conclusions}
In summary, we have adapted a linear sigma model used in the context of the study of the QCD axial anomaly, to an arbitrary number $N_f$ of equal mass fermions. 
Eqs. (\ref{eq:main}) provide simple formulas for the tree level spectrum. Using preliminary  \cite{latkmi17} and published \cite{Aoki:2016wnc} results, we found combinations of the masses that appear to vary slowly with the explicit chiral symmetry breaking and $N_f$. If confirmed by new numerical calculations at different masses and $N_f$, this would imply that the axial anomaly term 
plays a leading role in the mass spectrum and is essential to get a light $\sigma$.  The measurement of $M_{\etp}^2$ is a crucial ingredient to  locate the boundary of the conformal window.

The ratios in Figs. \ref{fig:rat} and \ref{fig:ratvrat} suggest to look for approximately $N_f$-independent relationships. It is expected \cite{dgrmp} that for $m_f$ 
large enough, the fermions decouple and the low energy theory is confining. Consequently, it is plausible that the massive theories at different 
$N_f$ have effective theories that can be smoothly connected. However, if $N_f=12$ and $N_f=8$ are on opposite sides of the conformal window, the massless limit of their effective theory 
should reveal important differences. Investigating the massless limit of $N_f=12$ within the framework proposed here should be quite interesting. Larger masses should also be investigated. If $R_\sigma$ stays approximately flat when $m_f$ increases, one would expect that the $\sigma$ remains the lightest state. On the lattice, at sufficiently large mass and coupling, the line of first order transition has an end point where one expects a 
second order phase transition and a light scalar in its vicinity \cite{Jin:2013hpa,zech1,zech2}. If this is a lattice artifact or something that could have 
a counterpart in the continuum is an open question that is worth investigating.

\acknowledgments
We thank Y. Aoki,  C. Bernard, T. DeGrand, D. Nogradi, E. Rinaldi, D. Schaich, B. Svetitsky and O. Witzel for useful conversations and suggestions.  
We thank E. Rinaldi for providing the preliminary $\etp$ masses \cite{latkmi17} with error bars.
This research was supported in part  by the Department of 
under Award Numbers DOE grant DE-SC0010113. 
This work started while attending the program ``Lattice Gauge Theory for the LHC and Beyond" at the  Kavli Institute for Theoretical Physics, which is supported by the National Science Foundation under Grant No. PHY11- 25915.
\appendix*
\section{Data for Figs. \ref{fig:rat} and \ref{fig:ratvrat}}
In this Appendix, we explain how we selected the data used in Figs. \ref{fig:rat} and \ref{fig:ratvrat}. 
We used Tables XVII, XXI, XXII, XXIII, XXVII, XXVIII and XXIX of \cite{Aoki:2016wnc} for $N_f=8$ and Table XXXVIII (with $\beta=4$) for $N_f=12$. 
We always used the largest volume available. For instance, $36^3\times 48$ for the four masses with $am_f=0.02$ and $N_f=8$. 
So different masses may have different volumes. 
The volume effects are typically a bit smaller than the quoted systematic and statistical errors but not negligible. The rest of the data comes from the graphs of Ref. \cite{latkmi17}  available on the Lattice 2017 link of Ref. \footnote{\url{https://makondo.ugr.es/event/0/session/96/contribution/350/material/slides/0.pdf}}. This is collected in  Table I. 
After our article was submitted, the data presented in Ref.  \cite{latkmi17}  appeared in a preprint  \cite{er17}. 
In addition, a  linear sigma model  (without anomaly term) discussed  for $N_f=8$ at the same conference appeared as a preprint \cite{ag17}.
\begin{table}[h]
  \begin{center}
    \begin{tabular}{|c|c|c|c|c|c|}
\hline    
$N_f$&$am_f$&$aM_\pi^2$&$aM_\sigma^2$&$aM_{a_0}^2$&$aM_{\etp}^2$\\
\hline
8&0.012 & 0.164(1) & 0.151(15) & 0.279(10) & 0.875(55)\\
8&0.015 & 0.186(1) & 0.162(23) & 0.310 (10)& 0.954(63) \\
8&0.020 & 0.221(1) & 0.190(17) & 0.365(11) & 0.956(49) \\
8&0.030 & 0.281(2) & 0.282(27) & 0.480(39) & 0.945(69) \\
8&0.040 & 0.335(2) & 0.365(43) & 0.567(23) & 0.977(42) \\
\hline
12&0.040 & 0.2718(7)& 0.24(1) & 0.3820(21)& 0.815(43)\\
12&0.050 & 0.3186(4) & 0.28(2) & 0.4469(25)& 0.850(44) \\
12&0.060 & 0.3629(3) & 0.30(2) & 0.5032(36)& 1.026(59) \\
\hline
 \end{tabular}
  \end{center}
  \caption{LatKMI data used for the graph.  }
  \label{fig:kmi}
  \end{table}

  \end{document}